# Automatic Diagnosis of Carotid Atherosclerosis Using a Portable Freehand 3D Ultrasound Imaging System

Jiawen Li, Yunqian Huang, Sheng Song, Hongbo Chen, Junni Shi, Duo Xu, Haibin Zhang, Man Chen*, Rui Zheng*

*Abstract*—The objective of this study is to develop a deep-learning based detection and diagnosis technique for carotid atherosclerosis using a portable freehand 3D ultrasound (US) imaging system. A total of 127 3D carotid artery scans were acquired using a portable 3D US system which consisted of a handheld US scanner and an electromagnetic tracking system. A U-Net segmentation network was firstly applied to extract the carotid artery on 2D transverse frame, then a novel 3D reconstruction algorithm using fast dot projection (FDP) method with position regularization was proposed to reconstruct the carotid artery volume. Furthermore, a convolutional neural network was used to classify healthy and diseased cases qualitatively. 3D volume analysis methods including longitudinal image acquisition and stenosis grade measurement were developed to obtain the clinical metrics quantitatively. The proposed system achieved sensitivity of 0.714, specificity of 0.851 and accuracy of 0.803 respectively for diagnosis of carotid atherosclerosis. The automatically measured stenosis grade illustrated good correlation (r=0.762) with the experienced expert measurement. The developed technique based on 3D US imaging can be applied to the automatic diagnosis of carotid atherosclerosis. The proposed deep-learning based technique was specially designed for a portable 3D freehand US system, which can provide more convenient carotid atherosclerosis examination and decrease the dependence on clinician's experience.

*Index Terms*—3D ultrasound imaging, automatic carotid atherosclerosis diagnosis, carotid artery segmentation, reconstruction with regularization.

## I. INTRODUCTION

CAROTID atherosclerosis is one of the major causes of stroke which is the world's second leading cause of death [1]. The prevalence of carotid atherosclerosis is 36.2% in Chinese people over 40 years old [2]. The pathological features of carotid atherosclerosis are increase of intima-media thickness (IMT) and appearance of atherosclerosis plaques. Magnetic resonance imaging (MRI), computed tomography angiography (CTA) and digital subtraction angiography (DSA) are several commonly used methods for visualizing and characterizing carotid artery features [3]-[7]. However, these methods still have some limitations during application due to invasiveness, ionizing radiation, portability etc.; and the approaches are very time-consuming and expensive which can't satisfy the need of large scale of examinations in different environments especially for community and rural areas. 2D ultrasound (US) as a non-invasive and low-cost method is widely used in the examination of carotid plaque. However, there are several disadvantages of traditional 2D US in the current examination of carotid atherosclerosis. Firstly, it is mainly carried out by experienced sonographers in hospital, and becomes a huge burden for health care system. Secondly, routine health check is difficult for carotid atherosclerosis patients especially in rural or undeveloped area. Thirdly, routine ultrasound examination is a tedious, laborious, experience-dependent work for sonographers. Lastly, some metrics such as intima-media thickness (IMT), plaque thickness, plaque area, usually assess the severity of the carotid atherosclerosis in 2D US images, which is prone to variability and lack of 3D morphology of carotid plaque [8]-[9]. Therefore, it is of great importance to develop a portable, reliable and cost-effective automatic ultrasound diagnostic technique for carotid atherosclerosis screening. 3D US carotid artery imaging approaches can provide plaque volume estimation, 3D morphology of plaque and other 3D metrics for carotid atherosclerosis diagnosis, which are found to be more accurate

This work was sponsored by Natural Science Foundation of China (NSFC) under Grant No.12074258. (Jiawen Li and Yunqian Huang are co-first authors.) (Corresponding authors: Rui Zheng, Man Chen.)

Jiawen Li, Sheng Song, Duo Xu and Haibin Zhang are with School of Information Science and Technology, ShanghaiTech University, Shanghai, China.

Hongbo Chen is with School of Information Science and Technology, ShanghaiTech University, Shanghai 201210, China, also with Shanghai Advanced Research Institute, Chinese Academy of Sciences, Shanghai 200050, China, and also with University of Chinese Academy of Sciences, Beijing 100049, China.

Yunqian Huang and Junni Shi are with Tongren Hospital, Shanghai Jiao Tong University School of Medicine, Shanghai, China.

Man Chen is with Tongren Hospital, Shanghai Jiao Tong University School of Medicine, Shanghai, China (e-mail: maggiech1221@126.com)

Dr. Rui Zheng is with School of Information Science and Technology, Shanghai Engineering Research Center of Energy Efficient and Custom AI IC, ShanghaiTech University, Shanghai, China (phone: 86 21-2068 4452, e-mail: zhengrui@shanghaitech.edu.cn)

to evaluate the progress of carotid atherosclerosis [10]-[13].

The 3D imaging system mainly includes mechanical scanning or tracked freehand scanning using various sensors, for example electromagnetic tracked senor, optical tracked sensor, etc. [14]-[15]. Compared to other tracking methods such as mechanical system, electromagnetic (EM) tracking system shows more flexibilities and penetration, which benefits 3D freehand US carotid artery imaging system. EM tracking system has been widely applied to surgical navigation. Harald *et al* [16] used various combinations of radiolucent bed inserts, EM tracking system and cone beam computerized tomography (CBCT) systems to register CBCT scans onto preoperative CT scan, and linked them with a 3D model to enable real-time surgical navigation. Gabrielle *et al* [17] demonstrated the feasibility of EM navigation for breast-conserving surgery by providing real-time position and distance information of the cautery device relative to the tumor, facilitating accurate navigation. For the EM tracking system, position and rotation of the US probe will be captured by electromagnetic sensor. Given the image and position information, reconstruction algorithm will be applied to obtain 3D US volume.

The automatic diagnosis of carotid atherosclerosis focuses on finding the biomarkers on ultrasound images, for example vessel wall area, vessel wall volume or total plaque volume, etc. [18]-[21]. These biomarkers are all bounded by the two boundaries of vessels, the media-adventitia boundary (MAB) and the lumen-intima boundary (LIB). Thus identifying these two boundaries is an important issue during the carotid atherosclerosis diagnosis. In recent years, deep learning methods have achieved excellent performance in medical image processing [22]-[24]. Jiang *et al.* designed a novel adaptive triple loss for carotid artery segmentation. To utilize 3D information in 3D volume of the carotid artery, Jiang *et al.* [25] introduced a fusion module to the U-Net segmentation network and yielded promising performance on carotid segmentation task. Zhou *et al.*[26] proposed a deep learning-based MAB and LIB segmentation method, and a dynamic convolutional neural network (CNN) was applied to image patches on every slice of the 3D US volume. LIB segmentation was performed by U-Net based on the mask of the MAB since the LIB is inside the MAB. The method achieved high accuracy but initial anchor points were still manually placed. Ruijter *et al.* [27] created a generalized method to segment LIB using CNN. Several U-Nets were compared and the experiments showed that the combination of various vessels such as radial, ulnar artery or cephalic vein, improved the segmentation performance of carotid artery. After segmentation, a 3D-geometry can be obtained for further assessment of therapy. Van Knippenberg *et al* [28] proposed an unsupervised learning method to solve the lack of data in carotid segmentation task. Azzopardi *et al.* [29] designed a novel geometrically constrained loss function and received improved segmentation results. Zhou *et al.*[30] proposed a voxel based 3D segmentation neural network to segment the MAB and LIB in 3D volume directly. Although the proposed algorithm achieved high accuracy with fast process, user's interaction is yet required to identify ROI in the first and last slices of the volume.

After region of interest (ROI) i.e., carotid artery is identified, further analysis needs to be performed to obtain significant clinical information for carotid atherosclerosis diagnosis such as the existence of plaque, carotid stenosis grade, type of the plaque, etc. Zhou *et al.*[31]-[32] applied 8 different backbone and UNet++ segmentation algorithm trained on 2D longitudinal US images to segment the plaque region and calculate the total plaque area. Xia *et al.* [33] employed a CNN to categorize segmented carotid images into normal cases, thickening vessel wall cases and plaque cases. Ma *et al.*[34] proposed a multilevel strip pooling-based convolutional neural network to investigate the echogenicity of plaque which was found to be closely correlated with the risk of stroke. Shen *et al.* [35] proposed a multi-task learning method, the authors combined ultrasound reports and plaque type labels to train a CNN to classify four different plaque type. Zhao *et al.* [36] utilized a novel vessel wall thickness mapping algorithm to evaluate therapeutical performance on carotid atherosclerosis. Zhou *et al.* [37] utilized the unsupervised pretrained parameters of U-Net to train a plaque segmentation network with a small 3D carotid artery ultrasound dataset. Saba *et al.* [38] used a deep learning based method to measure the carotid stenosis. Three systems based on deep learning were evaluated on 407 US dataset and achieved AUC of 0.90, 0.94 and 0.86 on the longitudinal US images, respectively. Biswas *et al.* [39] proposed a two-stage model for joint measurement of atherosclerotic wall thickness and plaque burden in longitudinal US images. The results showed that the proposed method achieved lower error compared to previous methods.

The current 3D carotid imaging device was mainly based on mechanical system and hard to transport, therefore the portable freehand 3D ultrasound imaging system was required which can be easily applied for various scenarios. However, for the freehand 3D ultrasound reconstruction, the requested small voxel size and various noise would lead to reconstruction artifacts[40]-[42]. On the other hand, the clinicians in different scenarios were usually inexperienced so that the diagnosis results might be inaccurate and hard to reproduce compared with sonographers in clinical ultrasound department. In this paper, we developed a new detection and classification technique based on deep-learning algorithms for carotid atherosclerosis diagnosis which can be employed to a portable freehand 3D US imaging system for fast screening. Compared to other methods mainly focusing on carotid vessel wall segmentation [24], [26], [27], [30], the proposed method aimed at exploring the automatic and experience-independent technique and framework for fast diagnosis.

The main contributions are outlined as follows. Firstly, an imaging and diagnosis framework including deep-learning based segmentation, 3D reconstruction and automatic volume analysis was developed for fast carotid atherosclerosis diagnosis using a portable freehand 3D US system. Secondly, a novel position regularization algorithm was designed to reduce the reconstruction artifacts caused by freehand scan. Lastly, post analysis including automatic stenosis measurement from 3D volume data provided qualitative and quantitative results for atherosclerosis diagnosis.

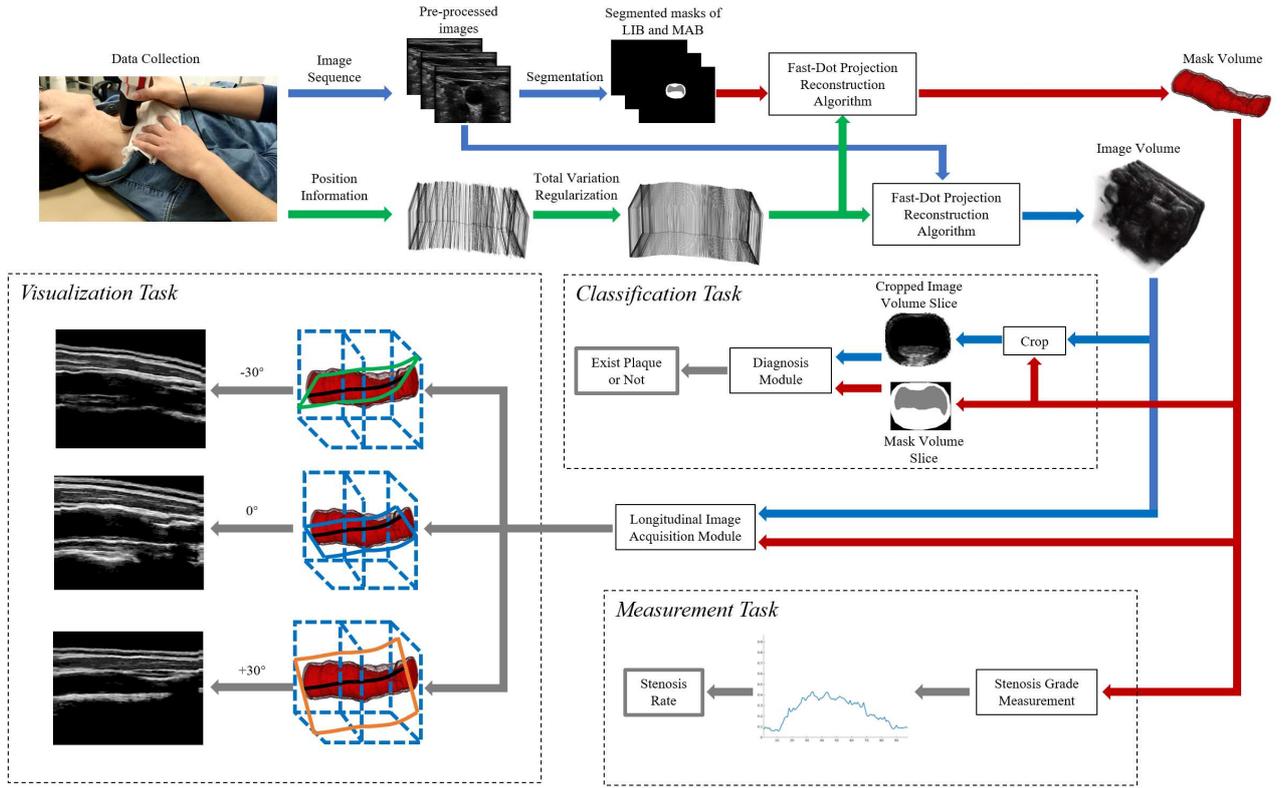

Fig. 1. The pipeline of the proposed technique and corresponding algorithm. The top row demonstrated the process of the data acquisition, extraction of ROI and 3D reconstruction. The bottom row represented the process of three diagnosis tasks based on the 3D volumes. The original image sequence and corresponding position information were firstly obtained by the 3D US device. U-Net segmentation algorithm and regularized Fast-Dot Projection algorithm were applied to extract the ROI and 3D carotid volume. Then the diagnosis tasks included automatic stenosis grade measurement, longitudinal image acquisition and healthy/diseased cases classification were conducted based on the reconstructed volume.

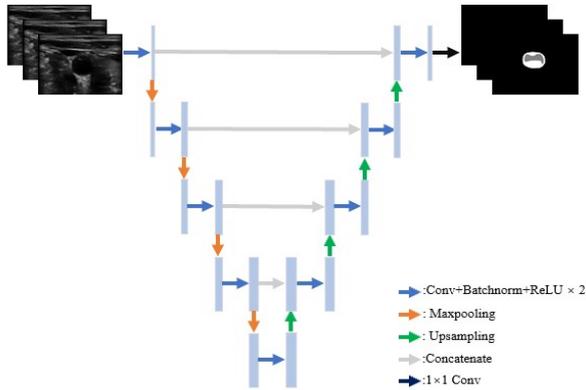

Fig. 2. The architecture of the segmentation module.

## II. METHODS

Fig. 1 showed the overview of data processing procedure including transverse image segmentation, 3D volume reconstruction, detection of carotid atherosclerosis and 3D carotid volume analysis. As the result, the image volume reconstructed from the original 2D image series and the mask volume reconstructed from the segmented carotid masks were acquired to complete three diagnosis tasks: 1) classification task: to diagnose whether existing plague or not; 2) measurement task: to measure the stenosis rate; and 3) visualization task: to achieve longitudinal images by cutting the

3D volume at different angles for better visualization of carotid artery. All subjects involved in this study consented to participate in this experiment, which was approved by the local ethics committee.

### A. MAB and LIB Segmentation

Three consecutive frames were concatenated in channel dimension which is proved to be useful to improve the segmentation accuracy [43].

Since the adjacent frames contained lots of redundant information, the pseudo labels were generated using pseudo-labeling method to reduce the work load [44]. One out of every 5 neighbor frames was selected to be manually labeled by experienced sonographers and the adjacent four frames were inferred to obtain the pseudo labels by the network which was trained using the labeled frames. All generated pseudo labels were checked visually, the labels would be corrected if the segmentation was incorrect.

U-Net was employed to segment the MAB and LIB for the transverse US image sequence [45]. The architecture of the network was illustrated in Fig. 2. The loss function of the segmentation module was the combination of DSC loss and cross-entropy loss.

### B. 3D Reconstruction with Regularization

After the MAB and LIB were segmented in every slice of US image sequence, the 3D volume of carotid artery was reconstructed using the Fast Dot Projection (FDP) method [46].

However, some disturbances caused by low precision of the electromagnetic sensor, inevitable hand shaking and breathing movement during scan, would lead to the reconstruction errors and artifacts. To improve the image quality and decrease the uncertainty of 3D reconstructed volume, a total variation regularization [47] method was integrated with FDP reconstruction algorithm.

(1) For all the position information obtained from 3D US device, it could be formulated as a set of rotation matrix $R$ and a translation $t$. The tuple $(\boldsymbol{R}, \boldsymbol{t})$ consisting of all $R$ and $t$ formed the special Euclidean group $SE(3)$ which was the semi-direct product of the rotation group $SO(3)$ and the translation group. Therefore, the $SE(3)$ can be formulated as:

$$SE(3) = \left\{ \begin{pmatrix} R & t \\ 0 & 1 \end{pmatrix} : R \in SO(3), t \in \mathbb{R}^3 \right\} \quad (1)$$

(2) The position signal obtained by the 3D US system could be considered as a set of entries which forms a vector $\boldsymbol{P} = (\boldsymbol{p_1}, \boldsymbol{p_2}, \dots, \boldsymbol{p_k}) \in M^k$, where $k$ was the number of entries and $k \in N$, $M^k$ was a manifold and $M = SE(3)$. Another signal set $\mathbf{x}$ was considered to be found when the following formula is minimal.

$$\mathrm{E}(\mathbf{x}) = \mathrm{D}(\mathbf{x}, \mathbf{p}) + \alpha \mathrm{R}(\mathbf{x}), \alpha > 0 \quad (2)$$

where $\mathrm{D}(\mathbf{x}, \mathbf{p})$ was the data term to limit the distance between original signal $\mathbf{p}$ and result signal $\mathbf{x}$. $\mathrm{R}(\mathbf{x})$ is a regularization term to penalize the position saltation in the signal $\mathbf{x}$.

(3) The deviation penalized term $\mathrm{D}(\mathbf{x}, \mathbf{p})$ could be defined as:

$$D(\mathbf{x}, \mathbf{p}) = \sum_{i=1}^{k} (h \circ d)(\mathbf{x}_i, \mathbf{p}_i) \quad (3)$$

Where $d(\mathbf{x}_i, \mathbf{p}_i)$ was the length of the geodesic which was defined as a shortest path on $M$ between two pose $\mathbf{p}$ and $\mathbf{q}$ [47]. $h$ was defined as following:

$$h(s) = \begin{cases} s^2, & s < 1/\sqrt{2} \\ \sqrt{2}s - 1/2, & \text{otherwise} \end{cases} \quad (4)$$

Which was the Huber-Norm.

(4) For the regularization term, it could be defined as the following:

$$R(\mathbf{x}) = \sum_{i=1}^{k-1} (h \circ d)(\mathbf{x}_i, \mathbf{x}_{i+1}) \quad (5)$$

where $d(\mathbf{x}_i, \mathbf{x}_{i+1})$ could be considered as the first-order forward difference. The optimize problem in (2) could be solved using a cyclic proximal point algorithm.

However, the original regularized algorithm couldn't handle the scanning positions with large backward movements. In this case, the position array was not sequential according to the coordinates, therefore pose re-rank algorithm was proposed. Concretely, considering the centroid point of every frame from the 2D segmented image sequence as $\boldsymbol{C_k} = (\boldsymbol{c_1}, \boldsymbol{c_2}, \dots, \boldsymbol{c_k})$, the PCA (principal components analysis) algorithm was conducted in $\boldsymbol{C_k}$ and a new matrix $\boldsymbol{D_k}$ was obtained. The first column of the matrix was the principal vector $v_k$, then a set of vectors $c_d$ could be acquired by projecting every centroid vector $c_k$ to $v_k$.

$$\boldsymbol{c_d} = \boldsymbol{c_k} - \frac{\boldsymbol{c_k} \cdot v_k}{v_k \cdot v_k} v_k \quad (6)$$

The new position sequence was finally obtained by sorting the L2-norm of the $\boldsymbol{c_d}$ set. The re-ranked position sequence was substituted into the total variance regularization algorithm to obtain the optimal positions.

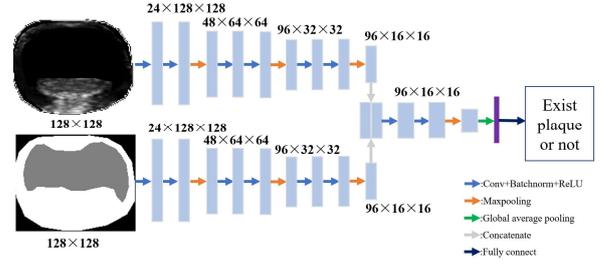

Fig. 3. The architecture of the diagnosis module.

### C. Carotid Atherosclerosis Diagnosis

The US scans including the segmented and reconstructed volume were classified into healthy case and carotid atherosclerosis case using a diagnosis network. As illustrated in Fig. 3, there were two inputs for the diagnosis module. It had been proved that the morphological information was helpful for the network to classify the normal or abnormal (diseased) images [48], therefore the mask of LIB and MAB extracted from each slice of the reconstructed volume was used as one input. The other input was the cropped ROI which was determined by the max bounding rectangular of the mask, and in the cropped image, the intensity in the region between LIB and MAB was set to the original value, while the intensity of region inside lumen and outside vessel wall was set to 0. For each input stream, it consisted of three repeated blocks, and each block consisted of two consequent basic convolutional sub-block and a max pooling layer. The basic convolutional sub-block was composed of a convolutional layer, a batch normalization module and a linear rectification unit. The number of channels for each repeated block was set to (24, 48, 96). The fusion block concatenated the high-level features of two streams and combined information by introducing a basic convolutional sub-block. After fusion block, the global average pooling (GAP) layers and a fully connected layer were applied to output the diagnosis result. We used focal loss in the diagnosis module.

The scan would be diagnosed as a carotid atherosclerosis case if any consecutive 5 transverse slices from the reconstructed volume were classified as plaque existing.

### D. 3D Carotid Volume Analysis

The clinical diagnostic parameters such as plaque thickness, plaque length, stenosis grade, plaque area, plaque type, etc. can be directly calculated from the reconstructed carotid artery volume. To validate accuracy of the proposed method, the longitudinal US images of carotid artery were obtained by cutting the volume in different angles, and the stenosis grade was calculated.

Stenosis rate is usually used to evaluate the stenosis grade. For the slices which are diagnosed as atherosclerosis, stenosis degree can be evaluated using the LIB and MAB masks. The diameter stenosis rate is usually calculated to evaluate the stenosis grade in clinic. We denote it as following equation which follows European Carotid Surgery Trial (ECST) stenosis grade measurement method [49].

$$S_{diameter} = \frac{L_{wall}}{L_{wall} + L_{lumen}} \quad (7)$$

where $L$ represents the length of respective area. The metric is ranged from 0 to 1, the greater number indicates the more severe stenosis. The length of vessel wall $L_{wall}$ and length of lumen $L_{lumen}$ are illustrated as Fig. 4. The diameter stenosis rate is the max $S_{diameter}$ which is calculated using all points in MAB boundary.

The longitudinal carotid US images are usually used to calculate plaque size and evaluate the morphology of plaque. Since the carotid artery is curved volume, the direct cutting along a fixed axis, i.e. directly extracting the 2D longitudinal images from 3D volume, may lead to missing of some structures. In this study, centroid points of carotid artery in transverse slices were selected to determine the cutting plane. Specifically, as illustrated in Fig. 5, denoting the centroid point of i-th slice in the volume as $C_i$, the line which was $\theta$ degree angled with y-axis through the centroid point $C_i$, was sampled as the i-th column of longitude image. In our experiment, the longitudinal US images were obtained by cutting the 3D carotid volume at the angles of 0°, ±15° and ±30°.

**Radially Sampling**

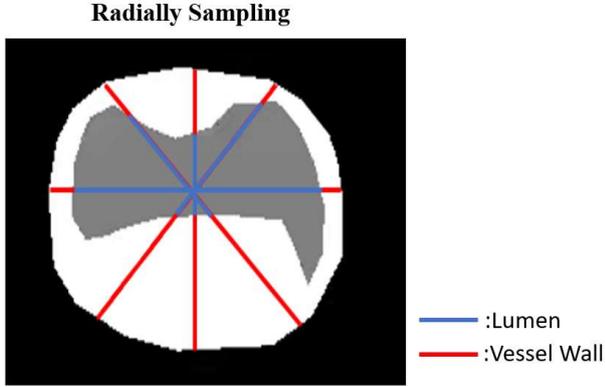

Fig. 4 The illustration of the approach to calculate the diameter stenosis.

## III. EXPERIMENTAL SETUP

### A. Data Acquisition and 3D Ultrasound Scan

A portable freehand 3D ultrasound imaging system was used to obtain 3D images of carotid artery (Fig. 6). The system included a 2D linear probe (Clarius, L738-K, Canada), an EM tracking system (Polhemus, G4 unit, U.S.A) and a host laptop computer (Intel i7-8700k CPU @ 3.70GHz, 32GB RAM) [50]. The 2D transverse US images were acquired by the probe while the corresponding position and orientation information were captured by the electromagnetic sensor. The images and orientation were acquired with a frame rate of 24 Hz.

During acquisition, the subjects took the supine position and were scanned (Fig. 6d), and the probe swept consistently along the long axis of common carotid artery from the proximal end to the distal end at the speed of approximate 10-15 seconds per scan. To reduce the reconstruction artifacts, fallback along the swept direction and large movement normal to the swept direction was maximally avoided. The inclusion criteria were based on visible plaques which were identified by expert. The stenosis grade larger than 70% was excluded from the dataset.

A total of 127 3D carotid artery scans from 83 subjects with stenosis range from 0% to 70% were obtained from local hospital. The age of the subjects was ranged from 51 to 86 years old (Male: 38, Female: 45). Each scan contained 122-250 2D transverse US images with image size of 640*480. 7036 2D images from 40 scans were manually delineated for MAB and LIB and labeled for healthy or diseased (with plaque) by experienced sonographers for further training of segmentation and classification network. The criterion of diseased was IMT lager than 1.5mm. All 127 scans were labeled for healthy or diseased (with plaque) by the same raters examining 2D images. In addition, stenosis grade and size of plaque (length and thickness) from randomly selected 20 scans were manually measured by expert using clinical 2D US device for verification of the proposed system and algorithm.

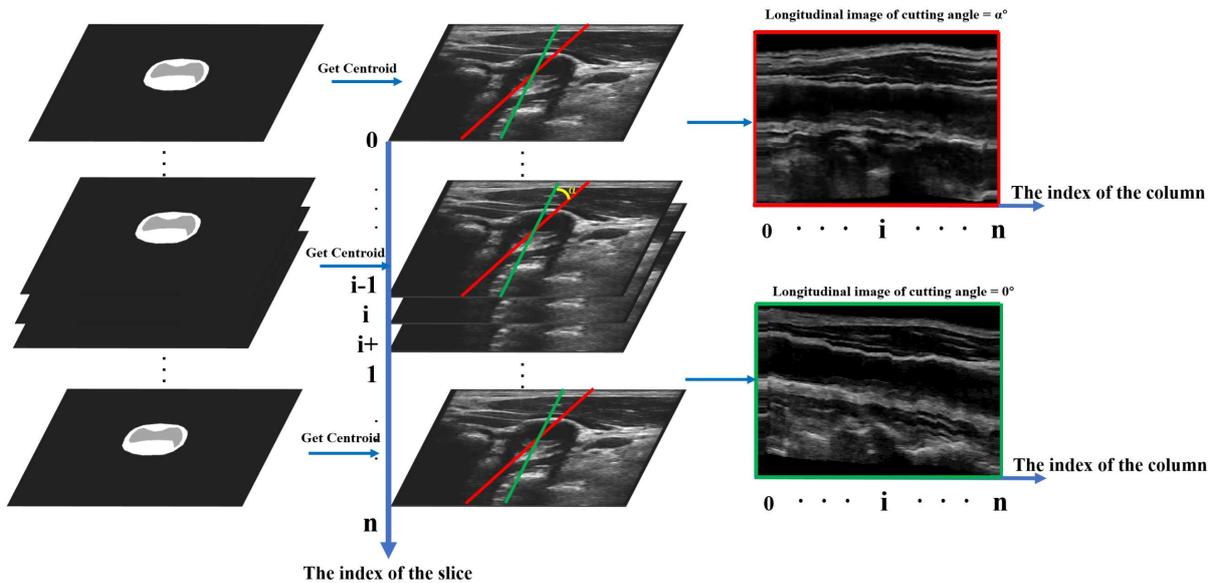

Fig. 5. The illustration of the cutting process. The centroid point was calculated by the segmented MAB mask for each slice in the volume. Then the line segment crosses the centroid point was set to conduct the cutting. The red and green line segments represent the different resample angles, respectively.

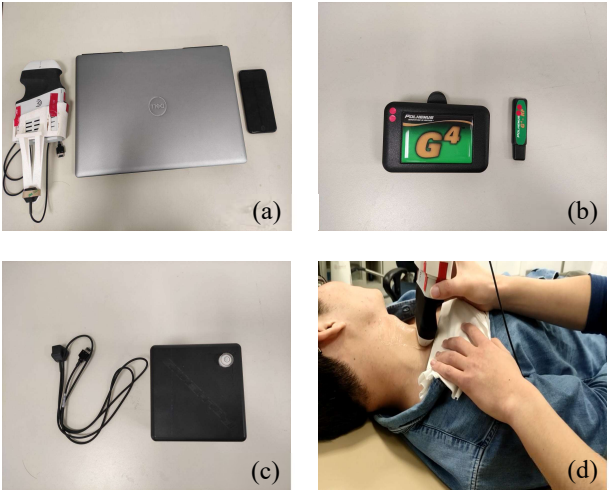

Fig. 6. Ultrasound scan using the freehand imaging system. (a) a handheld US scanner (left), a host laptop computer (middle) and an iPhone SE2 (right). (b) Tracking system including a hub (left) and a RF/USB module (right). (c) The sensor (left) and the magnetic source (right). (d) Ultrasound scan using the freehand imaging system.

## B. Training Methods

There were 25 randomly chosen scans (4546 2D images) for CNN training, 7 scans (1019 2D images) for validation and 8 scans (1471 2D images) for testing in order to build and verify the segmentation and diagnosis module. The original images were resized to 224*224. All training process were performed using Pytorch 1.5.1 and Python 3.7 on a NVIDIA RTX 4000 GPU. The two networks were trained separately. For the segmentation module, the applied data augmentation strategies including gamma transformation, rotation, zoom, horizontal and vertical flip, and Adam optimizer was used. The network was trained for 100 epochs with learning rate and batch size set to 0.005 and 8 respectively. For the diagnosis module, the image size was set to 128*128. Cropped and resized 2D US image segmented with the mask and the corresponding vessel wall mask were used for network training. Gamma transformation and horizontal & vertical flip were applied for data augmentation. The diagnosis network was trained for 50 epochs using Adam optimizer with learning rate and batch size set to 0.005 and 64 respectively.

## C. Diagnosis parameter measurement

To verify the regularized reconstruction and longitudinal images acquisition algorithm, 20 patients were recruited, and the longitudinal images from the 3D volumes reconstructed with and without regularization were compared with clinical images acquired by experienced sonographers visually, and the cutting angles were set as $\theta = -30°, -15°, 0°, 15°, 30°$.

The plaque length and thickness were manually measured on the 3D pseudo volume, the reconstructed 3D volume and the clinical images acquired by the experienced sonographers respectively, where 3D pseudo volume was acquired by directly stacking the 2D US image sequence according to the center position of each frame. The manual measurement of plaque length and thickness was conducted on the longitudinal images, while the cutting angle was chosen based on the carotid structural integrity and maximum stenosis grade. For plaque size measurement on longitudinal image of the reconstructed 3D volume, the pixel size was $0.2 \times 0.2 mm^2$. For the pseudo 3D volume, the velocity of the swept was assumed constant, therefore the pixel size of longitudinal image was determined by the distance of the swept which could be calculated by the magnetic sensor.

The stenosis grades for 20 clinical atherosclerosis patients were automatically measured on the longitudinal images from the reconstructed 3D volume and manually measured by experienced sonographers using clinical US device for the further comparison and analysis.

## D. Evaluation Metrics and Statistic Analysis

The dice similarity coefficient (DSC) and 95% Hausdorff distance (HD95) were used to evaluate the performance of the carotid sequence segmentation. DSC indicated the quantitative metric of the overlap region between the ground truth and prediction mask which was defined as follows:

$$DSC = \frac{2(P \cap L)}{P \cup L} \qquad (8)$$

where P, L were the prediction mask and ground truth. The Hausdorff distance was defined as Eq (9), which indicated the largest point-wise matching discrepancy:

$$HD(A, B) = max(hd(A, B), hd(B, A)) \qquad (9)$$

where

$$hd(A, B) = max_{a \in A}(min_{b \in B}||a - b||) \qquad (10)$$

$$hd(B, A) = max_{b \in B}(min_{a \in A}||b - a||) \qquad (11)$$

For the evaluation of the diagnosis module, the specificity, sensitivity and accuracy were calculated for both individual 2D US image and the whole scan.

The mean absolute difference (MAD) and standard deviation (SD) of plague size between the measurement results from the pseudo/reconstructed 3D volumes and the experienced sonographers were investigated. The stenosis grades were compared between automatic approach using the proposed technique and manual measurement using the clinical US device using the Pearson correlation analysis and MAD.

## IV. RESULTS

### A. Segmentation and Diagnosis Accuracy

Table I showed the average DSC and HD95 as well as the standard deviation between the human labels and prediction results. The comparison between 9 typical segmented images from U-Net and experienced sonographers was illustrated as Fig. 7, and the images were selected from different scans at some specific locations. Fig. 8 showed the frequency plot of the DSC and HD95 on the mean values of MAB and LIB. Table II showed the contingency table for the test set of 1471 2D transverse images. The sensitivity, specificity and accuracy were 0.73, 0.98 and 0.92 respectively. Table III showed the diagnostic results of carotid atherosclerosis for 102 US scans, and the sensitivity, specificity and accuracy of carotid atherosclerosis detection was 0.71, 0.85 and 0.80 respectively.

TABLE I. THE RESULTS OF VESSEL SEGMENTATION

| Metrics | category | |
|---|---|---|
| | *MAB* | *Lumen* |
| DSC | 95.67±2.76% | 94.17±5.60% |
| HD95(pixel) | 2.58±1.68 | 2.78±3.34 |

TABLE II. THE DETECTION RESULTS OF CA FOR INDIVIDUAL IMAGES

| Labels | Predictions | |
|---|---|---|
| | *Positive (plaque)* | *Negative* |
| Positive | 412 | 85 |
| Negative | 54 | 1024 |

TABLE III. THE DETECTION RESULTS OF CA FOR SCANS

| Labels | Predictions | |
|---|---|---|
| | *Positive (plaque)* | *Negative* |
| Positive | 25 | 10 |
| Negative | 10 | 57 |

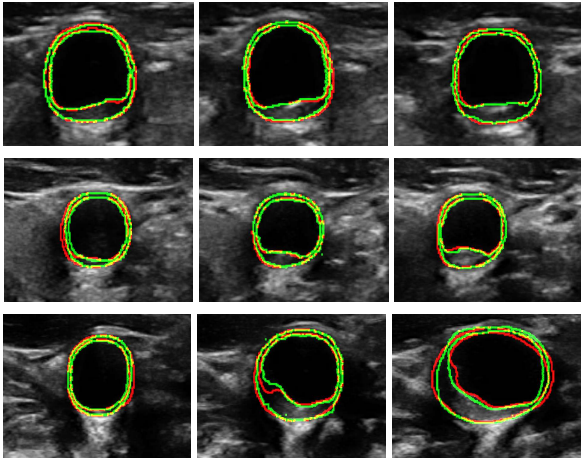

Fig. 7. Comparison of the auto segmentation from U-net (red) and manual segmentation from human labels (green).

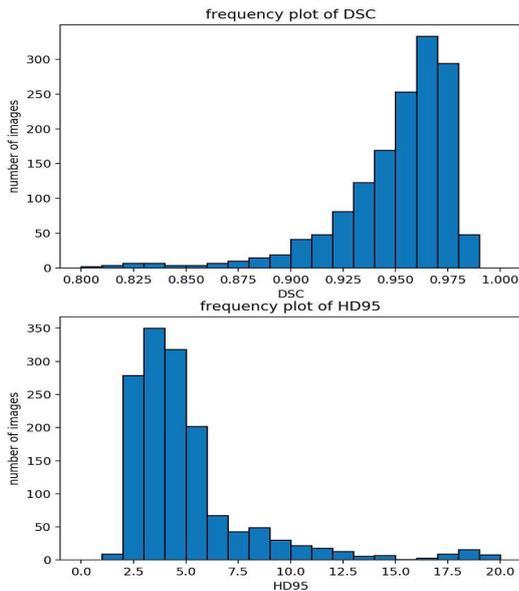

Fig. 8. The frequency plot of DSC and HD95 of mean of MAB and LIB.

## B. Reconstruction Accuracy

Fig. 9 demonstrated an example with large fallback of scanning trajectory. The results showed there were still noticeable artifacts (shown in the red boxes) when directly applying the regularized algorithm. The proposed re-rank algorithm could remove the reconstruction artifacts. Fig. 10 illustrated three representative examples of the longitudinal images reconstructed with or without regularization, and clinical images acquired by experienced sonographers. The results revealed that the regularized reconstructed volume was smoother with less image artifacts (Column 2).

Fig. 11 illustrated the 3D volumes reconstructed from the auto-segmentation and human labels respectively. The volumes were rendered by 3D-slicer (www.slicer.org). The results showed that the segmentation module achieved good agreement with human label. Furthermore, the sudden alteration of the lumen surface indicated the existence of the plaque (shown in black box).

Fig. 12 demonstrated comparison among 5 longitudinal images in different angles ($\theta = -30°, -15°, 0°, 15°, 30°$), the image directly projected to sagittal plane and the manually acquired image by expert from the same atherosclerosis patient. The results showed that the cutting images in different angles could reveal more structures of the carotid artery than the sagittal projection image. On the other hand, in Fig. 12, the image at 15° angle was most consistent with the clinical image obtained by expert using clinical US device, which indicated that the cutting longitudinal images from 3D volume could simulate the different scan angles operated by expert to locate the best observation view.

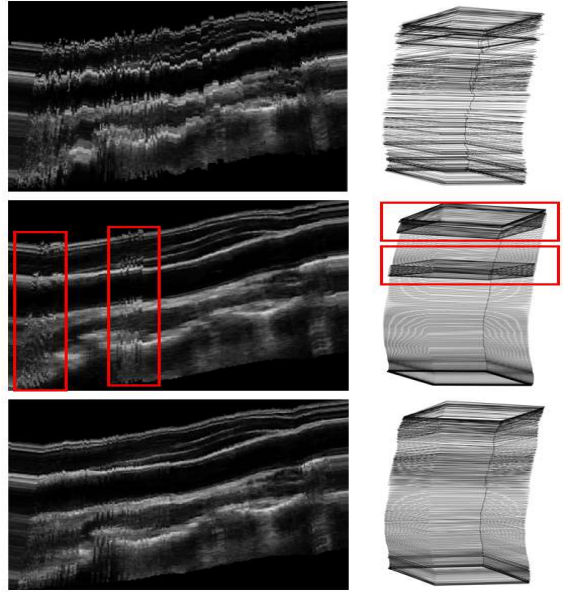

Fig. 9. Illustration of the proposed re-rank algorithm. The first row demonstrated the longitudinal image and corresponding position information without regularized algorithm; the second row represented the images which applied regularized algorithm; and the third row showed the images which used re-rank and regularized algorithm.

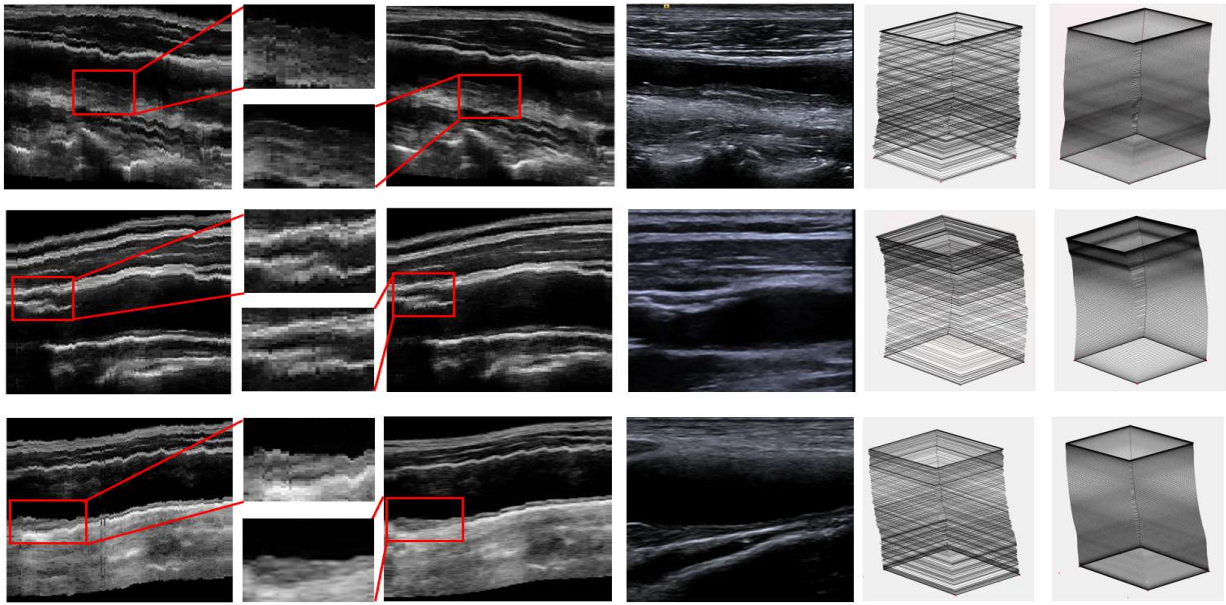

Fig. 10. Illustration of the US longitudinal images and the corresponding orientation information from three carotid atherosclerosis patients (by rows). **Column 1**: the image reconstructed without regularization algorithm; **Column 2**: the zoomed area for comparison between Column 1&3 which demonstrated the smoother results from the proposed algorithm; **Column 3**: the images reconstructed with regularization algorithm; **Column 4**: the images acquired by sonographers using clinical US devices; **Column 5**: the corresponding original position information; **Column 6**: the regularized position information.

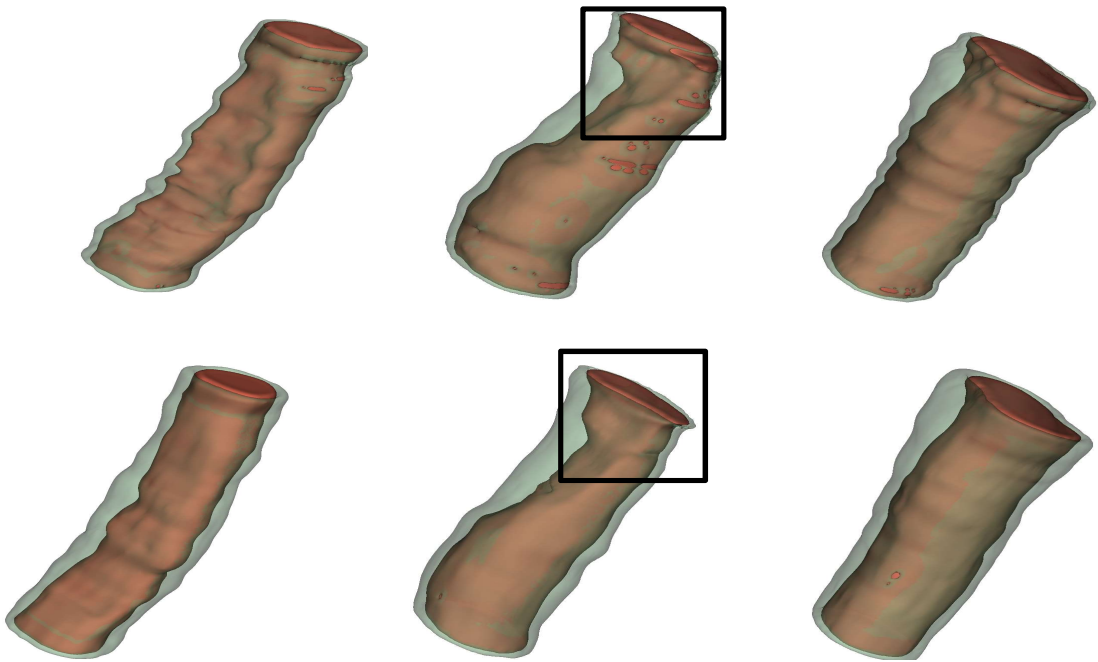

Fig. 11. The 3D volumes of three different subjects reconstructed from the auto-segmentation (the first row) and human labels (the second row). The translucent outer wall represents the vessel wall area, the inside red 3D volume represents the lumen area. The sudden alteration of the lumen surface indicated the existence of the plaque as black boxes show. The resolution of reconstruction is set to $0.2{\times}0.2{\times}0.2\ mm^3$.

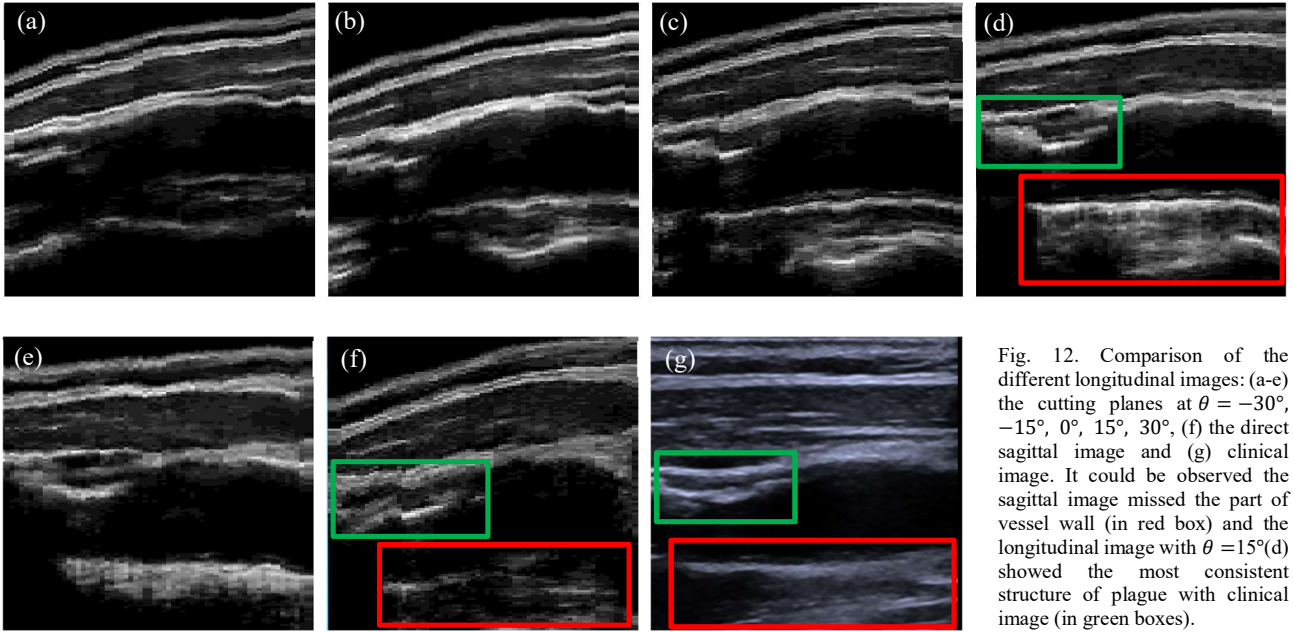

Fig. 12. Comparison of the different longitudinal images: (a-e) the cutting planes at $\theta = -30°$, $-15°$, $0°$, $15°$, $30°$, (f) the direct sagittal image and (g) clinical image. It could be observed the sagittal image missed the part of vessel wall (in red box) and the longitudinal image with $\theta = 15°$(d) showed the most consistent structure of plaque with clinical image (in green boxes).

TABLE IV. THE PLQUUE LENGTH/THICKNESS FROM THREE MEHTODS

| Patient ID | Expert Measurement (mm) | Direct Stacked Pseudo Volume (mm) | 3D Volume (mm) |
|---|---|---|---|
| 1 | 8.7/2.3 | 24.4/1.9 | 8.5/2.1 |
| 2 | 34.4/4.1 | 24.5/2.1 | 31.1/3.7 |
| 3 | 6.3/2 | 6.4/1.7 | 6.0/1.6 |
| 4 | 10.2/1.6 | 8.6/1.3 | 8.1/1.4 |
| 5 | 8.8/2 | 5.8/1.9 | 6.5/1.7 |
| 6 | 15.8/3.6 | 22.7/1.9 | 15.8/2.9 |
| 7 | 36/3.9 | 40.8/2.5 | 33.2/3.1 |
| 8 | 16.7/5.1 | 11.0/3.3 | 10.5/5.2 |
| 9 | 14.3/3.6 | 14.0/2.0 | 12.8/3.3 |
| 10 | 28.2/3.1 | 23.5/2.0 | 24.6/1.9 |
| 11 | 17.4/3.4 | 16.5/2.4 | 13.5/3.0 |
| 12 | 19/1.9 | 9.4/1.6 | 12.2/1.6 |
| 13 | 20.2/2.6 | 22.0/1.6 | 19.2/1.7 |
| 14 | 19/1.9 | 51.6/1.6 | 20.8/1.6 |
| 15 | 22.2/5 | 13.0/3.3 | 16.9/5.0 |
| 16 | 16/3 | 17.7/2.4 | 15.1/2.1 |
| 17 | 6.2/2.2 | 3.3/1.5 | 5.4/1.6 |
| 18 | 16.2/2.5 | 10.3/1.6 | 7.7/1.8 |
| 19 | 19.1/3.8 | 16.6/1.6 | 18.6/3.7 |
| 20 | 23.7/2.5 | 12.5/2.3 | 22.6/2.2 |

The plaque size (length and thickness) of 20 patients measured from the pseudo volume, reconstructed volume and images acquired by expert were shown in Table IV. Table V listed the measurement MAD between clinical US device and the two reconstruction methods. The results showed good agreement between the automatic measurement from the reconstructed volume and the manual method, while the plaque size measured by the pseudo volume showed large difference with the expert measurement. The results indicated that the 3D reconstruction could reveal true geometry and clinical metric of carotid artery.

TABLE V. MEASUREMENT MAD (N=20) BETWEEN CLINICAL US DEVICE AND THE TWO RECONSTRUCTION METHODS

| | Plaque Length (mm) | Plaque Thickness (mm) | Plaque Length (Relative Error) | Plaque Thickness (Relative Error) |
|---|---|---|---|---|
| Pseudo volume vs Expert | 6.54± 7.23 | 0.98± 0.65 | 40.0%± 48.0% | 29.4%± 14.0% |
| Reconstructed volume vs Expert | 2.65± 2.36 | 0.84± 0.62 | 15.4%± 13.6% | 26.0%± 13.2% |

*The MAD is calculated by sum of measurement errors divides N. The expert measurements are gold standard.

## C. Stenosis Measurement Accuracy

Fig. 13 demonstrated the linear correlation (r=0.762) of the stenosis grades automatically measured by the proposed US imaging system and manually measured by the experienced sonographers using clinical US device on 20 patients. The MAD of stenosis grade between automatic measurement and expert measurement was $8.24 \pm 4.32\%$. It indicated the proposed technique had the strong consistency with expert manual approach in carotid atherosclerosis diagnosis.

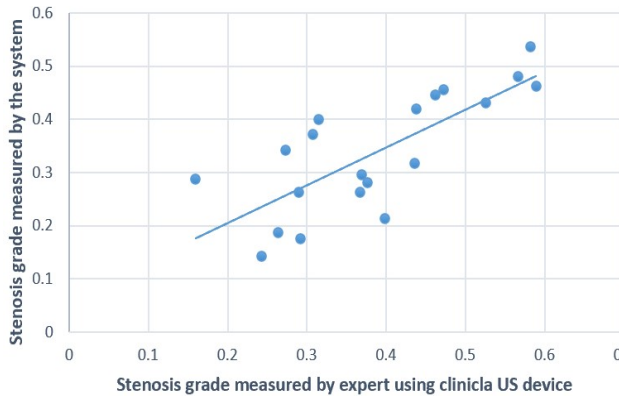

**Fig. 13.** Correlation of stenosis grade between the manual measurement by expert using the clinical US device and the automatic measurement from the proposed technique on 20 carotid atherosclerosis patients.

## V. DISCUSSION

In this study, we proposed a portable freehand 3D ultrasound imaging technique for carotid atherosclerosis diagnosis which could achieve real 3D geometry of carotid artery, and the diagnosis and measurement results from the proposed method showed good agreements with the results manually acquired by experienced clinicians. The system was transportable and less dependent on operator's experience, which makes it possible for routine health check in different environments such as community or rural area. In addition, the reconstructed geometry could provide 3D visualized carotid artery structure for further atherosclerosis evaluation.

Since the large position variation or fallback movement during scan would cause reconstruction artifacts, we designed a standard scan protocol for 3D carotid US data acquisition and analysis. The whole processing steps included 3D US data acquisition, MAB and LIB segmentation, 3D reconstruction, automatic classification and measurement. In practice, the intermediate results of each step could be reviewed and manually corrected by operator if necessary to ensure the accurate final results. The diagnosis result was based on two key points: the accurate segmentation of vessel area and the correct reconstruction volume. The segmentation determined the region of interest (ROI) used for the following analysis including automatic stenosis evaluation, plaque size measurement and 3D geometry visualization. The wrong masks might crop regions out of the carotid artery, mislead the diagnosis network and cause confusing diagnosis results. However, if the 3D volume was reconstructed directly from original 2D frames before segmentation, the reconstruction artifacts around MAB and LIB such as misplacement or severe blurring could lead to segmentation error of vessels, especially for some cases with large position variation (shown as Fig. 14). Therefore, we conducted segmentation on the original 2D US image sequence before 3D reconstruction for extracting the vessel area to reduce the influence of reconstruction artifacts.

For the reconstruction process, the failure reconstruction caused by large position variation could result in severe image artifacts which totally deviated the structure of the carotid artery as shown in Fig. 15a. For the freehand US scan, theoretically, the position information recorded by EM sensor would be consistent with US probe motion, i.e. the positions of US images should be sequential. However, the low precision of the sensor and inevitable hand jitter would lead to the noticeable measurement uncertainty of the position information along the scan direction, and moreover influence reconstruction accuracy. Therefore, we adopted a total variation regularization algorithm to smooth the track of the position information and decrease distortion and disconnection of image volume. The positions of the freehand scan can be regarded as continuous and sequential array; therefore the proposed regularization algorithm could reduce the uncertainty of position by minimizing a regularized function. Meanwhile, since even a small fallback movement during scan would invalidate the regularization algorithm, a re-rank strategy was designed to solve the unordered image sequence. The re-rank algorithm depended on segmentation results of 2D US image sequence. Thus, segmentation error would contribute to error in re-rank regularized reconstruction module. In the future, the reconstruction accuracy could be further improved using the neural network.

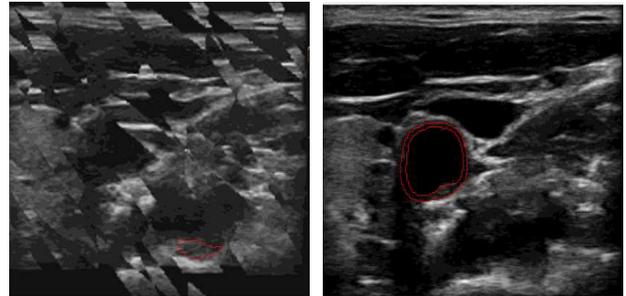

Fig. 14. Comparison of a transverse image at the same location from (left) the 3D volume directly reconstructed before segmentation and (right) original image sequence data. The severe artifacts on the left image led to wrong segmentation result (red circles).

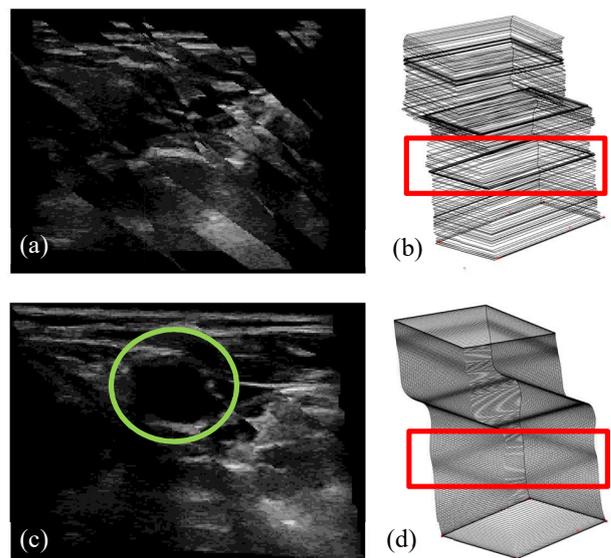

Fig. 15. Severe reconstruction artifacts caused by the large position variation. (a) a transverse image from the reconstructed volume (in red box in (b)) without regularization; (b) the orientation information without regularization, (c) a transverse image from the reconstructed volume (in red box in (d)) with regularization, and (d) the orientation information after regularization. The large distortion can be observed in (a), while in (c) the distortion was alleviated and clear vessel (in green circle) can be seen using the regularization algorithm.

After segmentation and reconstruction, the carotid artery volume could be obtained for further analysis such as healthy or diseased case diagnosis, plaque size measurement, plaque type identification and stenosis measurement etc. In the diagnosis module, the cropped and resized images instead of the whole US images were used as the input. Since the plaque was only located inside vessel wall area, removing useless information outside the vessel wall could accelerate network training and improve the detection accuracy. On the other hand, there may be low intensity area in the vessel region which could mislead the network and result in wrong classification since negative sample (no plaque) usually had low intensity in lumen area. Therefore, the MAB and LIB mask were introduced to combine the morphological information in original image to improve the detection accuracy. However, the proposed approach just utilized the consecutive 2D reconstructed transverse US images to detect plaque cases, thus some cases with small plaque size or severe artifacts were wrongly classified as no plaque. In the future, we will take the z axis information into account and use the whole 3D volume as input instead of detecting plaque by limited consecutive transverse slices to improve the accuracy of the diagnosis module.

We cut the carotid artery volume at different angles to acquire the longitudinal images, so that the clinical metric such plaque length and thickness could be directly measured from the 3D volume with no need of new acquisition in sagittal direction. The traditional clinical carotid artery US examination required appropriate positioning and angle between probe and neck, which greatly relies on the operator's experience to localize the plaque and obtain a high-quality US image. The proposed approach in our method was not only relatively convenient but could reveal the complete structure of the carotid artery with only one scan, and the longitudinal images obtained by our automatic method achieved great agreement with the images obtained by expert using clinical US device.

In segmentation module, we used U-Net to segment the MAB and LIB in 2D US image sequence. Every image in the sequence was treated as a single image for the segmentation network. However, this approach didn't exploit the context information in the adjacent frames. In addition, some cases with severe noise or shadowing would result in wrong segmentation as shown in Fig. 16. In the future, 3D convolution will be considered to correct the segmentation mistake by utilizing the context information of the adjacent frames, and sample size will be enlarged to improve the accuracy and robustness of the segmentation algorithm. More 3D metrics such total plaque area and volume, vessel wall volume would be evaluated for more accurate validation. On the other hand, the learning-based 3D reconstruction algorithm would be considered to improve the performance of reconstruction.

## VI. Conclusion

We have proposed an automatic 3D carotid artery imaging and diagnosis technique specially designed for the portable freehand ultrasound device. The technique applied a novel 3D reconstructed algorithm and a robust segmentation algorithm for automatic carotid atherosclerosis analysis. The results demonstrated that the technique achieved good agreement with manual expert examination on plaque diagnosis and stenosis grade measurement. The proposed technique showed high potential for fast carotid atherosclerosis examination and follow-ups, especially under the scenarios where professional medical devices and experienced clinicians are hard to acquire such as rural area or community with large population.

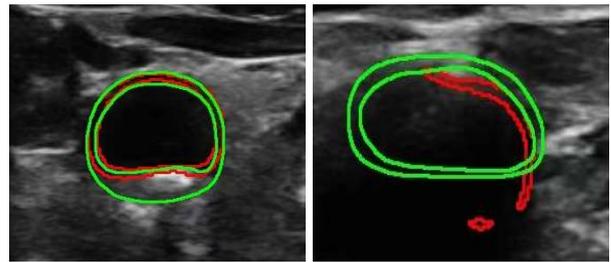

Fig. 16. Two typical wrong segmentation examples. The red lines show the automatic segmentation results by the segmentation module and the green lines show the human labels. The plaque was identified as the adventitia in the first case, and the vessel wall structure was disappeared in the second case.

## Acknowledgement

This work was sponsored by Natural Science Foundation of China (NSFC) under Grant No.12074258.